\documentclass[aps,prl,twocolumn,showpacs,superscriptaddress,groupedaddress]{revtex4}  
\usepackage{graphicx}  
\usepackage{dcolumn}   
\usepackage{bm}        
\usepackage{amssymb}   
\usepackage{amsmath}
\usepackage{float}
\usepackage[english]{babel}
\usepackage{amsmath}
\usepackage{amsfonts}
\usepackage{amssymb}
\usepackage{epsfig}
\usepackage{graphics,psfrag,rotating}
\usepackage{graphicx}
\usepackage{dcolumn}
\usepackage{bm}
\usepackage{epstopdf}
\usepackage{color}
\usepackage[usenames,dvipsnames,svgnames]{xcolor}
\usepackage{multirow}
\usepackage{float}

\usepackage{subfigure}

\usepackage{enumitem}

\def\3nab{\tilde{\nabla}}

\def\be {\begin{equation}}
\def\ee {\end{equation}}
\def\ba {\begin{align}}
\def\ea {\end{align}}

\def\bc {\begin{center}}
\def\ec {\end{center}}
\def\case#1/#2{\frac{#1}{#2}}

\newcommand{\bver}{\begin{verbatim}}
\newcommand{\ever}{\end{verbatim}}

\newcommand{\bea}{\begin{eqnarray}}
\newcommand{\eea}{\end{eqnarray}}
\newcommand{\beaa}{\begin{eqnarray*}}
\newcommand{\eeaa}{\end{eqnarray*}}

\def\case#1/#2{\textstyle\frac{#1}{#2}}

\RequirePackage[colorlinks,citecolor=blue,urlcolor=blue,linkcolor=blue]{hyperref}

\begin{document}


\title{Analysis of branon dark matter and extra-dimensional models with AMS-02}
\author{Jose A. R. Cembranos}
\affiliation{Departamento de F\'isica Te\'orica I and UPARCOS, Universidad Complutense de Madrid, E-28040 Madrid, Spain}
\author{\'Alvaro de la Cruz-Dombriz}
\affiliation{Cosmology and Gravity Group, Department of Mathematics and Applied Mathematics, University of Cape Town, Rondebosch 7701, Cape Town, South Africa}
\author{Peter K. S. Dunsby}
\affiliation{Cosmology and Gravity Group, Department of Mathematics and Applied Mathematics, University of Cape Town, Rondebosch 7701, Cape Town, South Africa}
\affiliation{South African Astronomical Observatory, Observatory 7925, Cape Town, South Africa}
\author{Miguel M\'endez-Isla}
\affiliation{Cosmology and Gravity Group, Department of Mathematics and Applied Mathematics, University of Cape Town, Rondebosch 7701, Cape Town, South Africa}

\date{\today}

\begin{abstract}
  In the context of  brane-world extra-dimensional theories,  we compute the positron production from branon dark matter annihilations and compare with the AMS-02 data.  Three different scenarios have been considered; the first two assume that either pulsars or dark matter may explain separately 
the whole positron fraction as measured by AMS-02, whereas the third one assumes that a suitable combination of these two contributions is needed. For all of them, exclusion diagrams for the brane mass and the tension of the brane, 
were obtained. Our analysis has been performed for a minimal, a medium and a maximum diffusion
model in one extra dimension for both pseudo-Isothermal and Navarro-Frenk-White dark matter
halos. 
Combined with
previous cosmological analyses and experimental data in colliders, constraints here enable us to set further bounds on the
parameter space of branons. In particular, in the case when pulsars fit the whole AMS-02 data, we have excluded mass-tension regions for masses and tensions smaller than $60.75$ TeV and $8.56$ TeV respectively. With regard to the scenario in which AMS-02 data are explained by a combination of dark matter and pulsars, masses and tensions smaller than $27.32$ TeV and $3.85$ TeV respectively turn out to be excluded. Finally, in the scenario with no pulsar contribution, a branon with a mass $38.1 \pm 0.2$ TeV and a tension $4.99 \pm 0.04$ TeV can fit well the experimental data.
\end{abstract}

\maketitle


\section{\label{sec:intro}1. Introduction}

The positron fraction when the electron/positron ($e^+/e^-$) theoretical background is only explained in terms of spallation of cosmic rays (secondary products) presents an excess for energies up to 10 GeV as 
measured by some detectors, such as AMS-02  \cite{Pizzolotto:2016lwk,Accardo:2014lma,Aguilar:2013qda}, PAMELA \cite{Adriani:2008zr,Adriani:2013uda} HEAT \cite{DuVernois:2001bb} or Fermi \cite{FermiLAT:2011ab}. 
This phenomenon has opened a wide discussion about the origin and the reliability of models of cosmic-ray propagation. 
In order to explain such an excess, it is necessary to introduce additional sources by injecting $e^+/e^-$ pairs. The main astrophysical sources \cite{DiMauro:2014iia,Yin:2013vaa} to interpret such a result are supernovae remnants (SNRs) \cite{Erlykin:2013xna}, the secondary production of positrons in the interstellar medium (ISM) generated by spallation of cosmic rays \cite{Delahaye:2008ua} and nearby pulsars \cite{Yuksel:2008rf}. Hence, taking into account the contributions from  averaged distant sources, fluxes from both local supernovae (Green Catalog \cite{Green:2009qf}) and pulsars from the ATNF database \cite{Taylor:1993ba} (such as Geminga, J1741-2054 or Monogem), the measurements of AMS-02 can be well fitted \cite{DiMauro:2014iia}.
Also, further explanations, such as annihilating dark matter (DM) in the Milky Way halo, have also been considered in the literature very recently  \cite{Feng:2017tnz,Laletin:2017qei,Cheng:2016slx,Carquin:2015uma}. Thus DM sources could either partially or completely explain the aforementioned excess  \cite{Delahaye:2007fr,Grefe:2011kh,Feng:2013zca,Ibarra:2013zia}. 
 Although some studies fit this excess with either astrophysical sources or DM separately, 
 the big space of parameters accounting for both astrophysical sources and DM contributions, enables us to describe the positron excess with a combination of them ({\it c.\,f.} \cite{DiMauro:2015jxa, Lin:2014vja,Lu:2015fdn} for recent attempts) bearing in mind that no source model can produce more $e^+/e^-$ than those observed by experimental data. 
 
 As is widely known, there is an outstanding variety of astrophysical and cosmological phenomena which require us to resort to DM  to obtain an accurate explanation. Among these observations, the most remarkable pieces of evidence are, among others, the presence of DM in the Coma \cite{Zwicky:1933gu,Zwicky:1937zza} and Bullet Clusters \cite{Clowe:2003tk,Clowe:2006eq}, the flat galactic rotation curves \cite{Rubin:1970zza,Rubin:1980zd}, gravitational lensing \cite{Massey:2010hh}, Nucleosynthesis abundances, the Cosmic Microwave Background (CMB) anisotropies and the growth of large structures \cite{Anderson:2013zyy}. From these pieces of evidence, some properties of DM can be inferred: namely, that it has to be non-relativistic 
at the moment of decoupling \cite{Kolb:1998kj}, be it stable or long-lived \cite{Audren:2014bca,Mambrini:2015sia}, effectively non-photon-interacting \cite{Wilkinson:2013kia}, collisionless \cite{MiraldaEscude:2000qt}, dissipationless \cite{Fan:2013tia}, smoothly distributed at cosmological scales \cite{LaceyOstriker1985} and sufficiently heavy \cite{Tremaine:1979we}. These evidences, together with the assumption of the particle nature of DM and the constraints of thermal decoupling, support one of the most suitable candidates for DM, the so-called Weakly Interactive Massive Particles (WIMPs). Several DM models have been proposed ({\it c.\,f.} \cite{Conrad:2017pms,Baltz:2004tj} and references therein) in order to explain the WIMPs features which cannot be accommodated within the Standard Model (SM) of elementary particles. 
In this paper, we shall focus on the so-called extra-dimensional brane-world theories \cite{Cembra2003,Kugo2001,AIPConf2003,astr0512569,Maroto2004,Cembranos0708.0235}, where a $4-$dimensional brane is embedded in a $D-$dimensional bulk ($D>4$). As a result of the brane fluctuations in the bulk, it is possible to define a pseudo-scalar Nambu-Goldstone boson, dubbed {\it branon}, 
which emerges due to an explicit translational symmetry breaking in the bulk space produced by the presence of the brane
 \cite{Maroto2004,Cembranos0708.0235,Sundrum085009,Bando3601,Dobado592,Cembranos026005,Alcaraz075010,Alcaraz096001}. 
Moreover, branon fields can be shown to be massive, stable and weakly interacting, what renders them competitive candidates for WIMPs which can naturally accommodate the correct amount and properties of DM particles.  
Although branons are prevented from decaying into SM particles by parity invariance on the brane, they may still annihilate by pairs into different channels of the SM. After the annihilation, these products  could decay or hadronise resulting in stable particles such as gamma rays, neutrinos, electrons-positrons or protons-antiprotons  \cite{Cirelli:2010xx}.

%
 Such particles then propagate from the DM halo through a convoluted transport process.
Such a propagation may cause 
signals to be directly detected at the Earth in the form of annihilation products or in the secondary processes of these stable particles with the galactic environment, such as radio signals of synchrotron radiation \cite{Fornengo:2011cn,Fornengo:2011iq,Natarajan:2015hma,McDaniel:2017ppt,Bull:2018lat} or gamma rays in the case of Inverse Compton Scattering.
As such, these signatures could potentially be measured by different detectors conforming the so-called DM indirect searches  \cite{Delahaye:2007fr,Grefe:2011kh,Feng:2013zca,Ibarra:2013zia}.

In this regard, 
cosmic rays from branon annihilations in different astrophysical sources have been studied thoroughly. For instance, gamma-rays analyses have been developed from observations of Cerenkov telescopes such as VERITAS, HESS, and MAGIC and satellites as Fermi \cite{branonsgamma, HESS,Ackermann:2012qk}. High-energy neutrinos have been studied from neutrino telescopes such as ANTARES or IceCube \cite{neutrinos, Fermani:2013fga}. Antiprotons have been analysed with
balloon experiments such as PAMELA, or satellites as AMS \cite{antipro}.
In fact, multimessenger astronomy study is a fundamental tool in the DM indirect searches realm since there are important uncertainties associated with simulations, backgrounds, diffusion and DM distributions \cite{MC}. Under standard assumptions, gamma rays turn out to provide the most constraining analysis \cite{branonsgamma, HESS}, however diverse cosmic rays might be able to constrain a different DM parameters region. For instance, the neutrino channel may be the most interesting one for studying very heavy DM \cite{neutrinos}. On the other hand, the positron analysis examined during the cause of this study, proves the local DM distribution and the shorter regime of diffusion models.

Indeed, taking into account the variety of sources that could explain the positron excess, in what follows we shall constrain the range of masses, the tension -and therefore the thermally averaged cross section - of the branons in order to ensure the DM contribution in the positron fraction is compatible with the AMS-02 results. 
At this stage, we draw our attention to the fact that eventual detection of indirect signals would not provide a conclusive evidence for DM since the uncertainties in the model-dependent DM interactions, DM density distribution in the halo and backgrounds from other astrophysical sources still remain entangled and are not fully understood yet. With this caveat in mind, this study focuses precisely on the possibility of the indirect detection method to obtain information about the nature of DM, abundance and properties using positron signals. 


The paper is organised as follows: in Section 2\ref{Sec:II}, we shall describe the propagation of positrons when obtained from the DM annihilation in our galaxy, the importance of each term in the  transport equation governing the propagation, 
and how the latter can be treated as a mere diffusion equation in the case of positrons
getting to the Earth at energies higher than 10 GeV. In addition, we shall discuss the source term that holds information about 
the DM model thermally averaged cross section, DM mass and its astrophysical disposition in halos. Then, in Section 3\ref{Sec:III}, in order 
to illustrate our line of reasoning, 
we shall 
present the rudiments of the underlying theory of  WIMPs, under consideration here, in the form of branons.
 Next, in Section 4\ref{Sec:IV} we shall summarise the technicalities emerging in the solution for the diffusion equation in terms of the Bessel-Fourier series. Such a solution would enable us to describe, in Section 5\ref{Sec:V}, the signature generated for the extra-dimensional branons in the positron fraction together with a 
 predefined background  model. Thus we shall provide our constraints for the branon parameter space. Finally, Section 6\ref{Sec:VI} shall be devoted to the main conclusions of this study. 

\section{\label{Sec:II}2. Transport of cosmic rays and the $e^+/e^-$ case} 
\subsection{\label{Sec:II.1}2.1 Generalities}

Cosmic rays are immersed in an environment governed by turbulent galactic magnetic fields. The departing point to tackle the propagation problem stems from the continuity equation 
\begin{equation}
\frac{\partial n}{\partial t}-D_{xx}{\bf \nabla} \cdot \left({\bf \nabla} n\right)=Q(\textbf{r},t)\,.
\label{Continuity_Eq}
\end{equation} 

This equation takes into account the number density of particles $n$ when there is a current of particles {\bf j} proportional to the concentration variation $\textbf{j}=-D_{xx} {\bf \nabla} n$.  where  $D_{xx}$ holds for  the diffusion coefficient \footnote{$D_{xx}$ is in principle a tensor \cite{Snodin:2015fza} which depends upon the cosmic rays energy and whose elements describe the diffusion when cosmic rays travel parallel or perpendicular to the magnetic field. 
}.
On the other hand, $Q(\textbf{r},t)$ 
holds for the source term which describes the injection of cosmic rays, in our case due to the DM annihilation. Using Eq. (\ref{Continuity_Eq}) as a first approximation to the problem it is then possible to add therein different mechanisms in the transport of cosmic rays along the galaxy \cite{Wu:2012ha} so the result is the so-called 
Ginzburg-Syrovatsky equation  \cite{Strong:2007nh,Yuan:2017ozr},

\begin{eqnarray}
&&\frac{\partial \psi}{\partial t}\,=\,{\bf \nabla} \cdot (D_{xx}{\bf \nabla}\psi-{\bf V}\psi)+\frac{\partial}{\partial p}D_{pp}\frac{\partial}{\partial p}\frac{1}{p^2}\psi \nonumber\\
&&-\frac{\partial}{\partial p }\left[b(p)\psi-(p/3)({\bf \nabla}\cdot{\bf {\rm \bf V}})\right]-\frac{1}{\tau_{f}}\psi-\frac{1}{\tau_{r}}\psi+Q(\textbf{r},p,t);\nonumber\\
&&
\label{Ginzburg-Syrovatsky_Eq}
\end{eqnarray}
%
being $\psi=n/E$ the number density of particles per unit of energy and  $p$ the total momentum of the particle at position $\bf r$.
The main contributions on the right-hand side of Eq. (\ref{Ginzburg-Syrovatsky_Eq})  can be summarised as follows:
 The first additional term  is characterised by the tensor $D_{pp}$ and can be understood as a diffusive process in the momentum space, the so-called reacceleration term \cite{Drury:2015zma,Drury:2016ubm}, which  considers the probability of having multiple accelerations of cosmic-ray particles in the interaction with the magnetohydrodinamic (MHD) shock wave in the interstellar medium (ISM). 

Secondly, another mechanism that could be relevant in the transport of cosmic rays is the convection associated with the galactic wind ${\rm \bf V}$, because of the stellar activity in late stellar stages that could push the ISM and the magnetic field out of the galactic plane, being the net effect an outflow perpendicular to the galactic plane. In addition, this mechanism not only implies a density redistribution but also a term of adiabatic losses $-\frac{p}{3} (\bf{\nabla}\cdot{\rm \bf V})$  in the expansion of the plasma \cite{Recchia:2016ylf}.

Then, the term in Eq. (\ref{Ginzburg-Syrovatsky_Eq}) referred to as radiative losses \cite{Sarazin:1999nz,Rybicki} is proportional to the rate of energy loss per unit of time $b(p)$.  This term might be of great importance for some cosmic-ray species, mainly because of its relevance in the dynamic of cosmic rays along the galaxy. Eq. (\ref{b(E)_contributions}) below will provide further details for the various contributions included in $b(p)$.
Finally, revisiting the Eq. (\ref{Ginzburg-Syrovatsky_Eq}) the strike of primary cosmic rays, i.e., those coming directly from the source, with ISM particles may produce secondary cosmic rays in a process dubbed spallation through fragmentation or radioactive decay of the particles being $\tau_{f}$ and $\tau_{r}$ the time scale for each instance respectively.

Now that we have thoroughly described the terms in Eq. (\ref{Ginzburg-Syrovatsky_Eq}), its form can be simplified by assuming the validity of a
quasi-linear theory regime, in which there is a magnetic field with short fluctuations $\left(\delta B\ll B\right)$, the diffusion tensor turns out to be a scalar $D_{xx}\sim D(R)$, where $R=pc/Ze$ is the rigidity that gives a particle response under a magnetic field. As a consequence, the most conventional diffusion model renders $D(R)=K_{0} \left(\frac{p}{m c}\right) \left(R/ {\rm GV}\right)^{\delta}$ with the rigidity $R$ measured in gigavolts. As we can see, the diffusion coefficient is dependent on the energy and can be parameterised by two constants, $K_{0}$ and $\delta$. 
Moreover, under this approximation 
the space diffusion coefficient $D_{xx}$ and the reacceleration parameter $D_{pp}$ are related by \cite{Thornbury:2014bwa,Drury:2015zma};
\begin{eqnarray}
D_{xx}D_{pp}=\frac{p^{2}\,V_{A}^{2}}{\delta\left(4-\delta\right)\left(4-\delta^{2}\right)}\,,
\label{Correlation_Dxx_Dpp}
\end{eqnarray}
where $V_{A} \sim 20$ ${\rm km/s}$ is the Alfv\'en velocity of the MHD wave. In the case of positrons, since they travel with velocities close to the speed of light, the diffusion term is one of the most relevant terms. The reacceleration parameter $D_{pp}$  is inversely proportional to the diffusion one according to Eq. (\ref{Correlation_Dxx_Dpp}). If the latter is dominant the probability of having a second acceleration can be ignored. On the other hand, the convection velocity takes values of $\left|{\rm \bf V}\right|\sim 10$ ${\rm km/s}$ \cite{Donato:2001ms} which is negligible when the spectra of positrons at the Earth is greater than $10$ GeV, the range of energies that we consider in this study.
In addition, herein we shall only calculate the contribution of positrons due to the DM annihilation without considering the spallation terms $-\frac{1}{\tau_{f}}\psi-\frac{1}{\tau_{r}}\psi$ in Eq. (\ref{Ginzburg-Syrovatsky_Eq})
since this mechanism is considered part of a background model which will be analised in Section 5\ref{Sec:V}. 
Once a steady-state has been reached, it is possible to obtain a purely diffusion equation for positrons from Eq. (\ref{Ginzburg-Syrovatsky_Eq}) yielding
\begin{eqnarray}
-K_{0}\left(\frac{E}{1\,{\rm GeV}}\right)^{\delta}\nabla^{2}\psi-\frac{\partial}{\partial E}(b(E)\psi)=Q(\textbf{r},E)
\label{Diff_Eq_positrons}
\end{eqnarray} 
where $b(E)$ includes the radiative losses in the energy space which can be split in the following contributions \cite{Sarazin:1999nz} 
\begin{eqnarray}
b(E)&=&b_{{\rm brem}}(E)+b_{{\rm Coul}}(E)+b_{{\rm ion}}(E)\nonumber\\
&&+\,b_{{\rm ISRF}}(E)+b_{{\rm syn}}(E)
\label{b(E)_contributions}
\end{eqnarray}
%
including
bremsstrahlung, Coulombian interactions, ionization of the medium, Inverse Compton Scattering  of the interstellar Radiation Field (ISRF) and synchrotron emission. In Table \ref{Table0} we have provided analytical expressions for each term in Eq. (\ref{b(E)_contributions}).
%
\begin{table}
\centering
\caption{Analytical expressions for each term in Eq. (\ref{b(E)_contributions}). $n_{e}$ is the number density of electrons in the plasma and $\gamma$ is the Lorentz factor, $U_{{\rm rad}}=0.9$ ${\rm eV}/{\rm cm}^{3}$ is the energy density of radiation (starlight, emission from dust and CMB), $\sigma_{T}$ is the Thomson cross section, $m_e$ is the electron mass, $c$ is the speed of light, $q_{e}$ is the electron charge, $n_{H}$ is the number density of the ionised neutral hydrogen, $I$ is the energy of ionization of neutral hydrogen
and $U_{B}=\frac{B^2}{8\pi}$ is the magnetic energy density in cgs.}
\label{Table0}
\begin{tabular*}{\columnwidth}{@{\extracolsep{\fill}}l@{}}
  \hline
  Radiative losses contributions   \\
  \hline
  $b_{\rm brem}(E)\simeq 1.51\cdot10^{-16}n_e\gamma\left[\rm{ln}(\gamma)+0.36\right]$     \\
  $b_{{\rm coul}}(E)\simeq 1.2\cdot10^{-12}n_e\left[1+\frac{{\rm ln}\left(\frac{\gamma}{n_e}\right)}{75}\right]$    \\
  $b_{{\rm ion}}(E)=\frac{q_{e}^{4} n_{H}}{8 \pi \epsilon_{0}^{2} m_{e}^{2} c^{3}\sqrt{1-\frac{1}{\gamma^{2}}}}[\ln \frac{\gamma\left(\gamma^{2}-1\right)}{2\left(\frac{I}{m_{e} c^{2}}\right)^{2}} -\,\left(\frac{2}{\gamma}-\frac{1}{\gamma^{2}}\right)\ln 2$ \\
	\hspace{1.5cm}$+\frac{9}{8\gamma^{2}}+\frac{1}{8}-\frac{1}{4\gamma} ] $\\
  \\
	$b_{{\rm ISRF}}(E)=\frac{4}{3} \frac{\sigma_{T}}{m_{e} c} \gamma^2 U_{{\rm rad}}$,\hspace{1.5cm}  $b_{{\rm syn}}(E)=\frac{4}{3} \frac{\sigma_{T}}{m_{e} c}  \gamma^2 U_B$       \\
	\hline
\end{tabular*}
\end{table}
%
%
%
 As such, at high energies 
 electrons/positrons lose energy 
mainly through Inverse Compton Scattering and synchrotron emission. Consequently, both $b_{{\rm coul}}(E)$ and $b_{{\rm brem}}(E)$ can safely be neglected in Eq. (\ref{Diff_Eq_positrons}) of our analysis.


%

As mentioned above, the $Q(\textbf{r},E)$ source term in Eq. (\ref{Diff_Eq_positrons}) includes information on the source injecting positrons in the environment \cite{Ibarra:2013cra,Bertone:2010zza}. Provided the only source of positrons is the DM annihilation, then 
\begin{eqnarray}
 Q(\textbf{r},E)=\frac{1}{2}\left\langle \sigma v\right\rangle\left(\frac{\rho(\textbf{r})}{M}\right)^2\sum_{i}\beta_{j}\frac{{\rm d}N_e^j}{{\rm d}E},
 \label{Q}
\end{eqnarray}     
where $\left\langle \sigma v\right\rangle$ is the total thermally averaged cross section of annihilation, $\rho(\textbf{r})$ is the DM density profile of the halo, $M_{\rm{}}$ is the DM mass, $\frac{{\rm d}N_e^j}{{\rm d}E}$ is the injection spectra of the positron due to DM annihilation at the annihilation point and $\beta_j= \frac{\left\langle \sigma v_{j}\right\rangle}{\left\langle \sigma v\right\rangle}$  the branching ratios providing the annihilation probability in one particular channel $j$.  Both the injection spectra 
and the thermally averaged cross section 
are DM model-dependent quantities, the latter will be analysed in Section 3\ref{Sec:III}. 

On the other hand, it is necessary to describe how DM is disposed in halos through DM density profile $\rho(\textbf{r})$. 
In the following  we shall consider both the (pseudo-) Isothermal DM halo profile \footnote{$\rho_{{\rm ISO}}(r)=\frac{\rho_0 r^{2}_{a}}{(r^{2}+{r^{2}_{a}})}$\,,with $r_{a}=5$ kpc and $\rho_0= 1.53$ GeV ${\rm cm}^{-3}$.}  
and a piecewise modified Navarro-Frenk-White (NFW) model \footnote{$\rho^{*}_{\rm NFW}(r)=\rho_{\rm NFW}(r_{0})\\
\times\sqrt{1+8.11\cdot{\rm sinc}\left(\frac{\pi r}{r_{0}}\right) 
+ 6.11\cdot\, {\rm sinc}\left(\frac{2\pi r}{r_{0}}\right)},  r<r_{0} \\
\rho_{{\rm NFW}}(r)=\frac{\rho_s}{\frac{r}{r_s}\left(1+\frac{r}{r_s}\right)^2}\,,\,\;\;r>r_{0}
\label{NFW_core}$ where $\rho_{s}=\frac{3{H^2({\rm z})}\delta_c}{8\pi G}$ contains information about the Universe at the redshift ${\rm z}$ when the halo collapsed and $r_0$ value represents the overlapping radius between both expressions.  For values of $r_{0}\approx 10^{-7}$ pc, we can ensure that solutions of the diffusion equation (\ref{Diff_Eq_positrons}) do not vary significantly at $r<r_0$.
} which are able to simulate the core at the centre of the galaxy according to the study in \cite{Delahaye:2007fr} and avoid the discontinuity at $r=0$ from which the usual NFW DM density profile is prone to. 

%
%
\section{\label{Sec:III}3. Branons as WIMPS candidates}

%

As mentioned in the Introduction, the fluctuations of the brane can be parameterised by branon fields. Such fields
from the point of view of an observer within the brane,  behave as WIMPs \cite{Cembranos:2007ve, Cembranos:2003mr,Kugo:1999mf,Cembranos:2003aw,Cembranos:2004fm,Cembranos:2004pj,Cembranos:2004sa,Cembranos:2004rg,Cembranos:2004eb,Cembranos:2005mu,Cembranos:2005im,Cembranos:2006mj,Maroto:2003gm,Maroto:2004qb,Cembranos:2008kg}. 
Moreover, in the context of low energy-effective Lagrangian theories, branons couple through the stress-energy tensor 
with an interaction suppressed by $f^4$, $f$ being the brane tension \cite{Sundrum:1998sj,Bando:1999di,Dobado:2000gr,Cembranos:2001rp,Cembranos:2001my,Alcaraz:2002iu,Cembranos:2004jp}, 
and thanks to the brane parity invariance, decay into SM particles is prevented.
Also, the branons annihilation cross sections depend solely upon the branon mass $M$, and the mass and spin of the SM particle  \cite{Cembranos:2003fu,Cembranos:2006mj}. 
For non-relativistic branons, the leading term in the
 thermally averaged cross section of annihilation into
Dirac fermions $\psi$ with mass $m_\psi$,
 becomes
\begin{eqnarray}
\langle \sigma_{\psi} v\rangle\,=\,\frac{M^2 m_\psi^2}{16\pi^2f^8}\left(M^2-m_\psi^2\right)\,\sqrt{1-\frac{m_\psi^2}{M^2}}\,,
\label{cross_section_fermion}
\end{eqnarray}
whereas for a massive gauge fields ($W$ or $Z$), of mass $m_{W,Z}$, it reads
\begin{eqnarray}
\langle \sigma_{W,Z} v\rangle\,&=& \,
\frac{M^2}{64\pi^2f^8}\,\left( 4\,M^4 - 4\,M^2\,{m_{W,Z}}^2 + 3\,
{m_{W,Z}}^4 \right)\nonumber\\
&&
\,\times{\sqrt{1 - \frac{{m_{W,Z}}^2}{M^2}}}\,,
\label{cross_section_WZ}
\end{eqnarray}
for a massless gauge field $\gamma$, the leading order is zero
\begin{eqnarray}
\langle \sigma_{\gamma} v\rangle&=&0\;,
\label{cross_section_massless}
\end{eqnarray}

\begin{figure}
\includegraphics[scale=0.7]{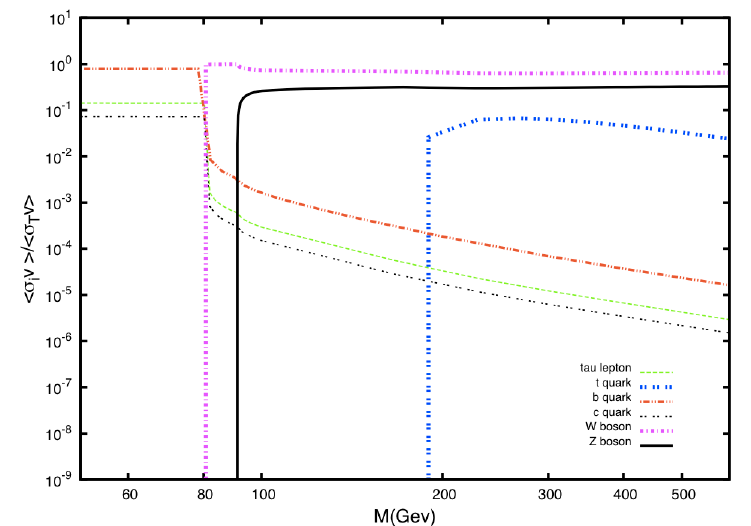}
\caption{\label{Fig_BR_branons}Annihilation branching ratios for extra-dimensional branons in different channels as taken from \cite{Cembranos:2012xp}. The annihilation into leptons is highly suppressed excepting for lighter branons (until a mass of $\sim 80$ GeV), although the annihilation is mainly via $b\bar{b}$. From this mass, other kinds of processes are opened, in which the annihilation is mostly via $Z\bar{Z}$ and $W^{+}W^{-}$. Our work is centered in this latter range of masses.}
\end{figure}

and, finally, for a (complex) scalar field $\Phi$ of mass $m_\Phi$:
\begin{eqnarray}
\langle \sigma_{\Phi} v\rangle\,=\,\frac{M^2}{32\pi^2f^8}\,{\left( 2\,M^2 + {m_\Phi}^2 \right) }^2\,
{\sqrt{1 - \frac{{m_\Phi}^2}{M^2}}}
\,.
\label{cross_section_complex}
\end{eqnarray}
These thermally averaged annihilations are represented in Figure \ref{Fig_BR_branons}
in all the allowed SM particles channels, i.e., fermions, vector gauge bosons and scalars. 
%
%
Such annihilation cross sections 
have been used to provide both
constraints and prospects on the brane-world theories parameters from tree-level processes
for colliders such as ILC, LHC or CLIC \cite{Alcaraz:2002iu,Cembranos:2004jp,Achard:2004ds,Creminelli:2000gh,Cembranos:2005jc}. Also, further astrophysical and cosmological bounds for brane-world theories were obtained in \cite{Cembranos:2007ve, Cembranos:2003mr,Kugo:1999mf,Cembranos:2003aw,Cembranos:2004fm,Cembranos:2004pj,Cembranos:2004sa,Cembranos:2004rg,Cembranos:2004eb,Cembranos:2005mu,Cembranos:2005im,Cembranos:2006mj,Maroto:2003gm,Maroto:2004qb,Cembranos:2008kg}. 

Since branons are described  by an effective field theory, their  phenomenology is qualitatively different from
other DM candidates described by renormalisable couplings, such as the neutralino
in R-parity conserved supersymmetric theories. In particular, leading branon couplings with SM
particles are provided by dimension 8 operators. Accordingly, in the non-relativistic limit, this implies a very 
strong dependence on the annihilation cross sections with the branon mass. This is a very distinctive
feature of these DM candidates which are then able to saturate the unitarity limit for freeze-out thermal
production at large brane tension scales \cite{Cembra2003,Kugo2001}.

\begin{figure}
\includegraphics[scale=0.33]{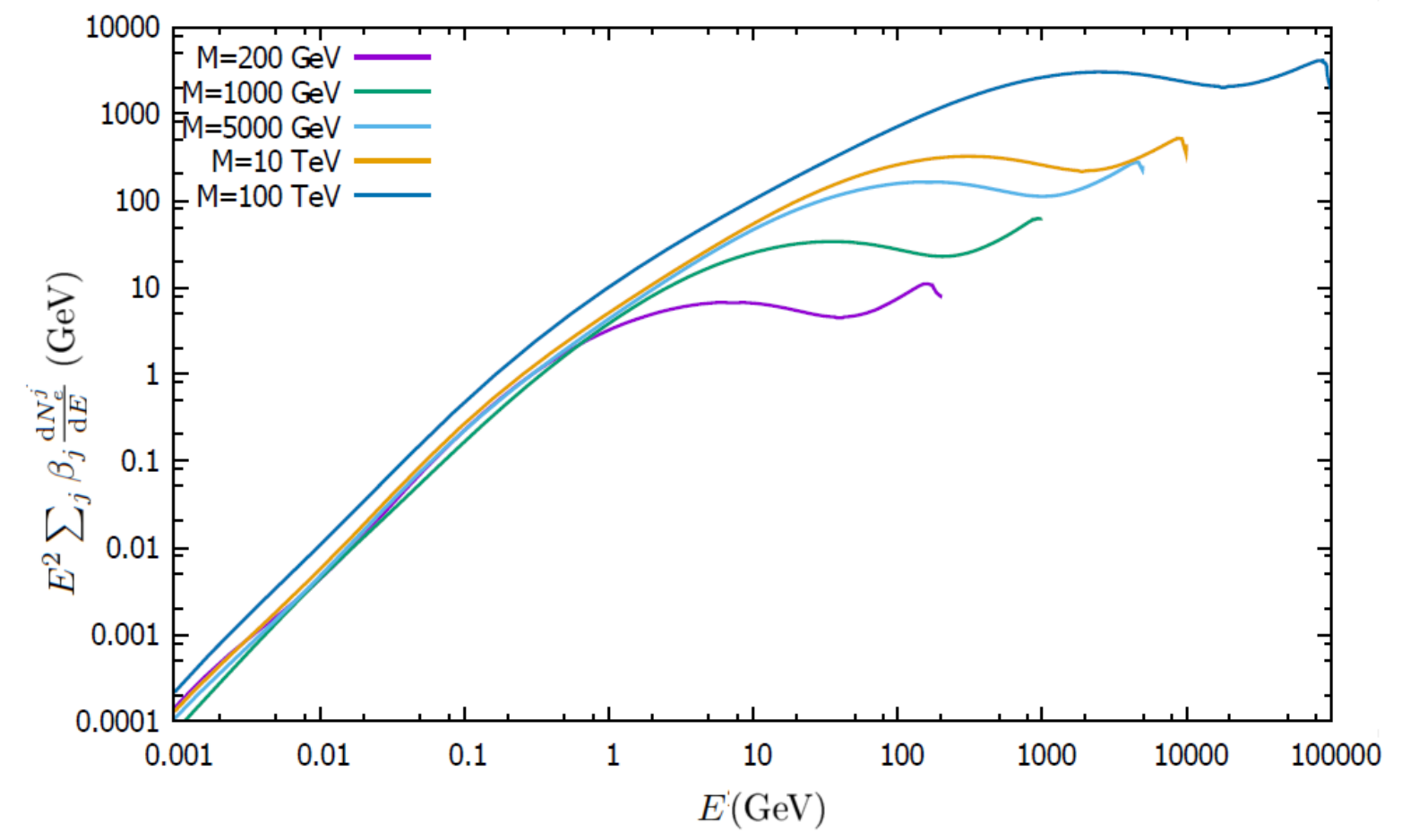}
\caption{\label{Fig_Branon_injection} The spectra of positron at the point of injection at the galactic environment (before propagation). Branching ratios are independent from the tension of the brane. As it is possible to see, the annihilation produces positrons mostly at high energies. In order to compute the total injection spectra we use the functions as provided by ${\rm PPC4DMID}$ \cite{Cirelli:2010xx,Ciafaloni:2010ti}}
\end{figure}

Hence, positrons are injected into the environment by branons annihilations are related to $\sum_{j}\beta_{j}\frac{{\rm d}N_e^j}{{\rm d}E}$ appearing in Eq.~(\ref{Q}). 
Figure~\ref{Fig_Branon_injection} demonstrates how the amount of positrons increases with the mass of the branon, 
so positrons, obtained from such  annihilations, mainly occur at high energies. However, after propagating one does not
expect to detect positrons at very high energies, since energy losses are more pronounced at this range. 

\section{\label{Sec:IV}4. Fluxes at the Earth}

The transport diffusion zone for positrons in our galaxy is modeled as a cylinder of radius $R_{g}$ and thickness $2 L_{z}$ centered at the center of the galaxy. At the coordinates $R_{g}$ and $L_{z}$, it is considered that the density of electrons/positrons is negligible with regard to the rest of the density in the galaxy, so that boundary conditions to solve the Eq. (\ref{Diff_Eq_positrons}) are $\psi\left(R_{g},z\right)=0$ and $\psi\left(r,\pm L_{z}\right)=0$. Then, the solution of the Eq. (\ref{Diff_Eq_positrons}) 
can be expressed in terms of the Bessel-Fourier series \cite{Delahaye:2007fr,Cirelli:2010xx} as follows,
 \begin{eqnarray}
 \psi(\vec{x},E)=\sum^{\infty}_{i=1}\sum^{\infty}_{n=0} J_{0}\left(\frac{\alpha_{0,i}}{R}r\right)\varphi_{b}\left(z\right)P_{i,n}(E),
 \label{Bessel}
\end{eqnarray}
where $J_{0}$  are Bessel functions \footnote{In our numerical resolution of Eq. (\ref{Diff_Eq_positrons}) 
for a truncated solution (\ref{Bessel}), including  $\{n,i\}$ terms, we have checked that the relative error when adding the $\left(n+1\right)^{th}$ or $\left(i+1\right)^{th}$ term does not exceed more than $0.1 \% $ with respect to the truncation at $\{n,i\}$
terms in the sum above. 
In the case of Isothermal profile with MIN diffusion  $i,n \simeq 100$ terms were required, whereas for NFW with MED diffusion it was necessary  $i,n \simeq 200$ terms and  for the MAX diffusion it was necessary around $i,n \simeq 400$ to obtain the required precision.} of the first kind with $\alpha_{0,i}$ zeros and $\varphi_{b}\left(z\right)=\sin\left(\frac{n \pi}{2L}(z+L_{z})\right)$. 
On the other hand, the dependence of the density $\psi$ with the energy is given by 
\begin{eqnarray}
%
P_{i,n}(E)&=&\frac{1}{b(E)}\int^{M_{{\rm }}}_{E}{\rm d}E_{s}\, Q_{i,n}(E_{s})\\
&&\times\exp\left\{-\frac{\lambda_{D}^{2}(E,E_{s})}{4}\left
[\left(\frac{n\pi}{2L_{z}}\right)^{2} + \left(\frac{\alpha_{i}^2}{R^2}\right)\right]\right\}.\nonumber
\label{psi_density_dependence}
\end{eqnarray}
%
In fact, the $e^+/e^-$ density solution (\ref{Bessel}) is a particular case of a more general solution based on the Green's method. In this method the cosmic-ray particles are generated at given spacetime coordinates 
with an energy $E_{s}$. Once the particles are injected at the source, they reach different coordinates 
with an energy $E$. The distance between these two points is called the diffusion length $\lambda_{D}$, which satisfies
\begin{eqnarray}
\lambda_{D}^{2}(E,E_{s})=4\int^{Es}_{E}{\rm d}
\varepsilon \,\frac{K_{0}\,\varepsilon^{\delta}}{b(\varepsilon)}.
\end{eqnarray}
Finally the factor $Q_{i,n}(E_{s})$ in expression (\ref{psi_density_dependence}) corresponds to the Bessel and Fourier transforms of the source term as follows
\begin{eqnarray}
Q_{i,n}(E_{s})&=&
\frac{2}{L_z\,R^{2}\,J_{1}^{2}\left(\alpha_{i}\right)}\\
&&\times\int^{R}_{0} \int^{L_{z}}_{-L_{z}} r\,{\rm d}r \,{\rm d}z\,\, J_{0}\left(\frac{\alpha_{0,i}}{R}r\right)\varphi_{b}\left(z\right) Q(\textbf{r},E_{s}).\nonumber
\end{eqnarray}
%
By considering that the Solar System is about $r\simeq8$ kpc and $z\simeq0$, we can obtain the fluxes of ${\rm electrons}/{\rm positrons}$ at the Earth as being purely generated by DM annihilation, i.e., without considering any contribution 
from standard astrophysical sources. Hence the positron flux at the Earth $\textbf{r}_{\odot}$ becomes
\begin{eqnarray}
\Phi^{{\rm DM}}_{e^{\pm}}(\textbf{r}_{\odot},E)=\frac{v_{e}\left(E\right)}{4\pi c} \psi(\textbf{r}_{\odot},E)\,,
\label{Positron_DM_flux}
\end{eqnarray}
where $v_{e}$ is the $e^+/e^-$ velocity which generally depends on the energy $E$. In our case we consider ultra-relativistic electrons/positrons, i.e., $v_e \left(E\right) \simeq c$. 
Figure \ref{Fig_flux_positrons} illustrates how after the propagation, the energy maximum for cosmic rays  does not happen at high energies, as occurred at the injection point (see Figure \ref{Fig_Branon_injection}). 
In the case of the branons with a tension $f=200$ GeV the maximum of $E$ is $~ 1 \%$ of the highest energy and in the case of $f=100$ GeV around $10^{-5} \%$. In addition, Figure \ref{Fig_flux_positrons}
also shows how the higher brane tensions $f$ are, the lower the received positron flux 
becomes.

\section{\label{Sec:V}5. Background model and comparison with AMS-02 results}

\begin{figure}[tbp]
\centering 
\includegraphics[width=.475\textwidth, height=6.0cm]{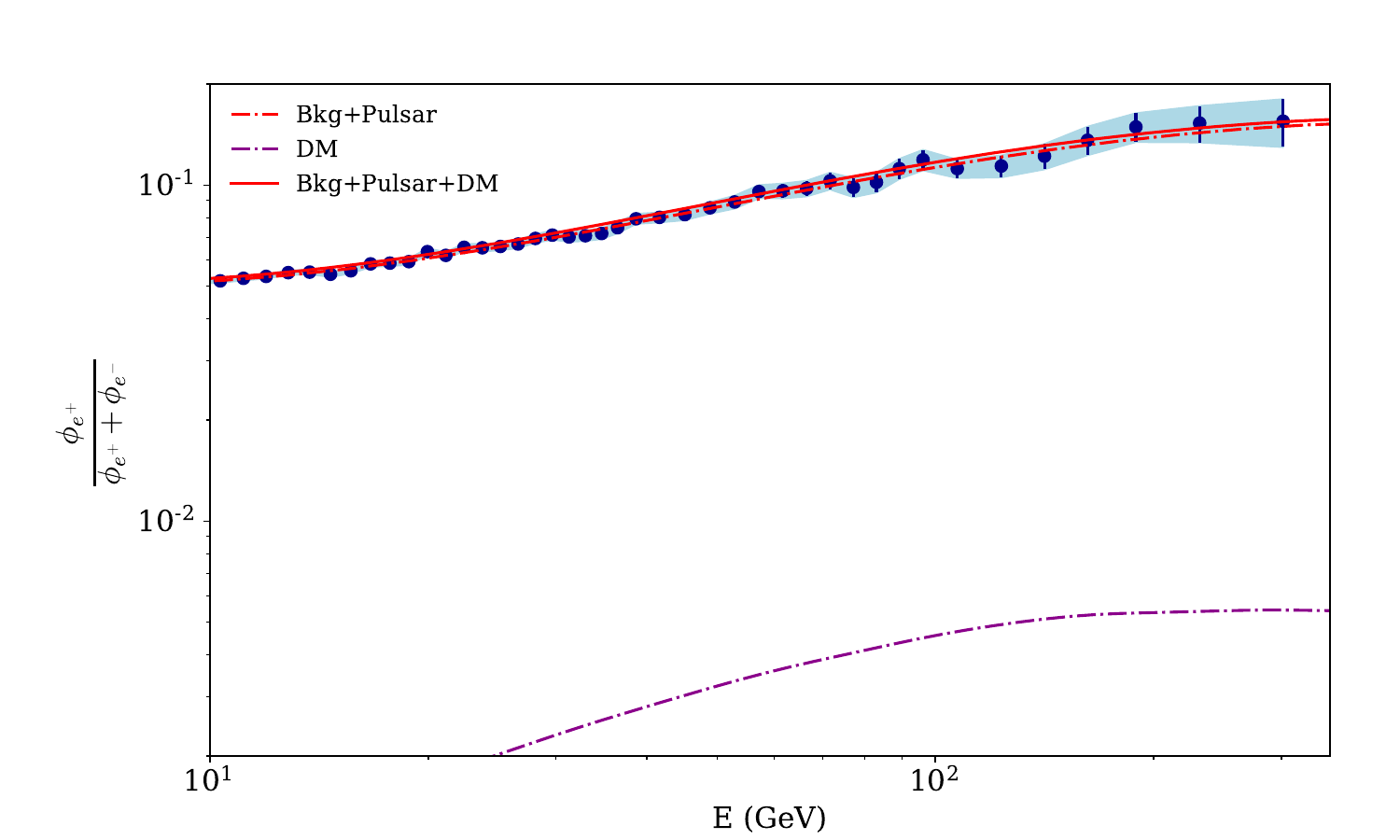}
\hfill
\includegraphics[width=.475\textwidth,height=6.0cm]{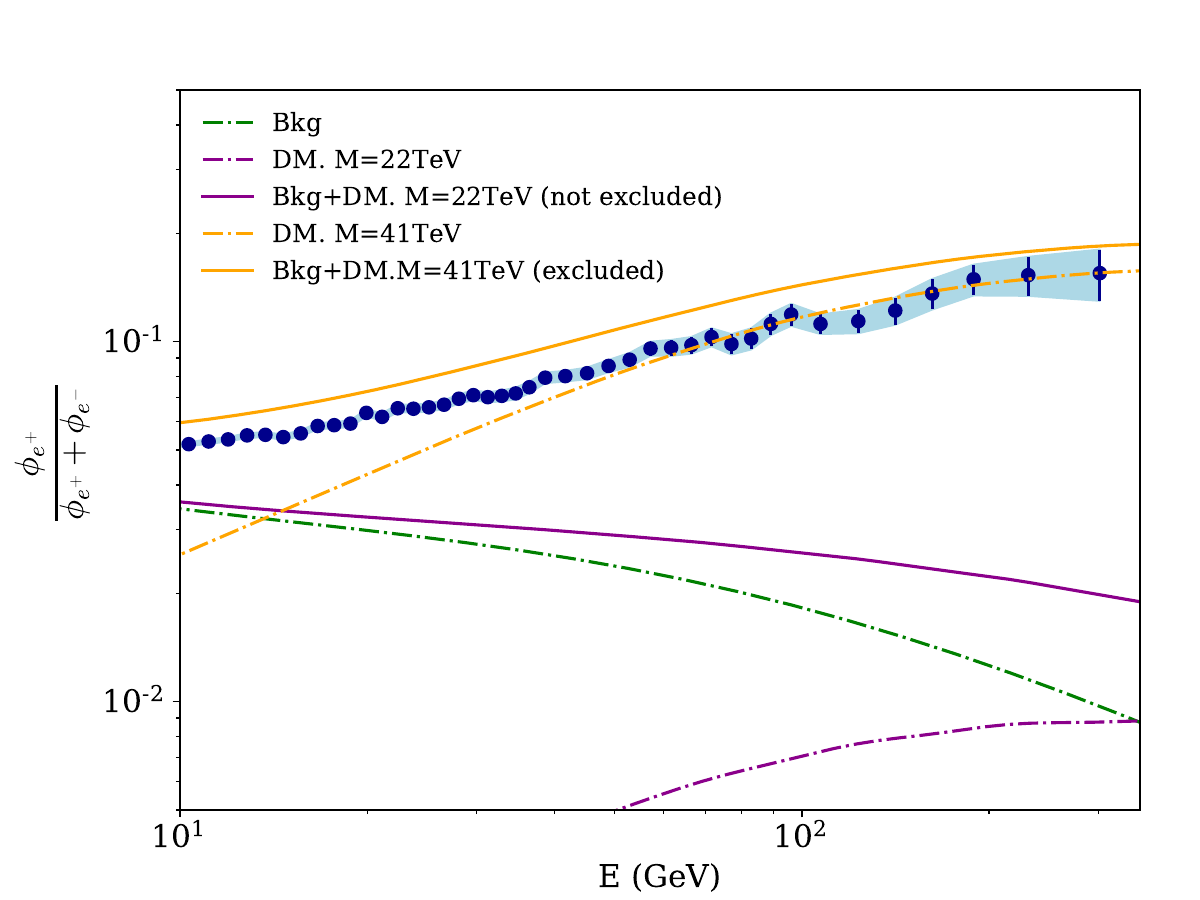}
\hfill
\includegraphics[width=.475\textwidth,height=6.0cm]{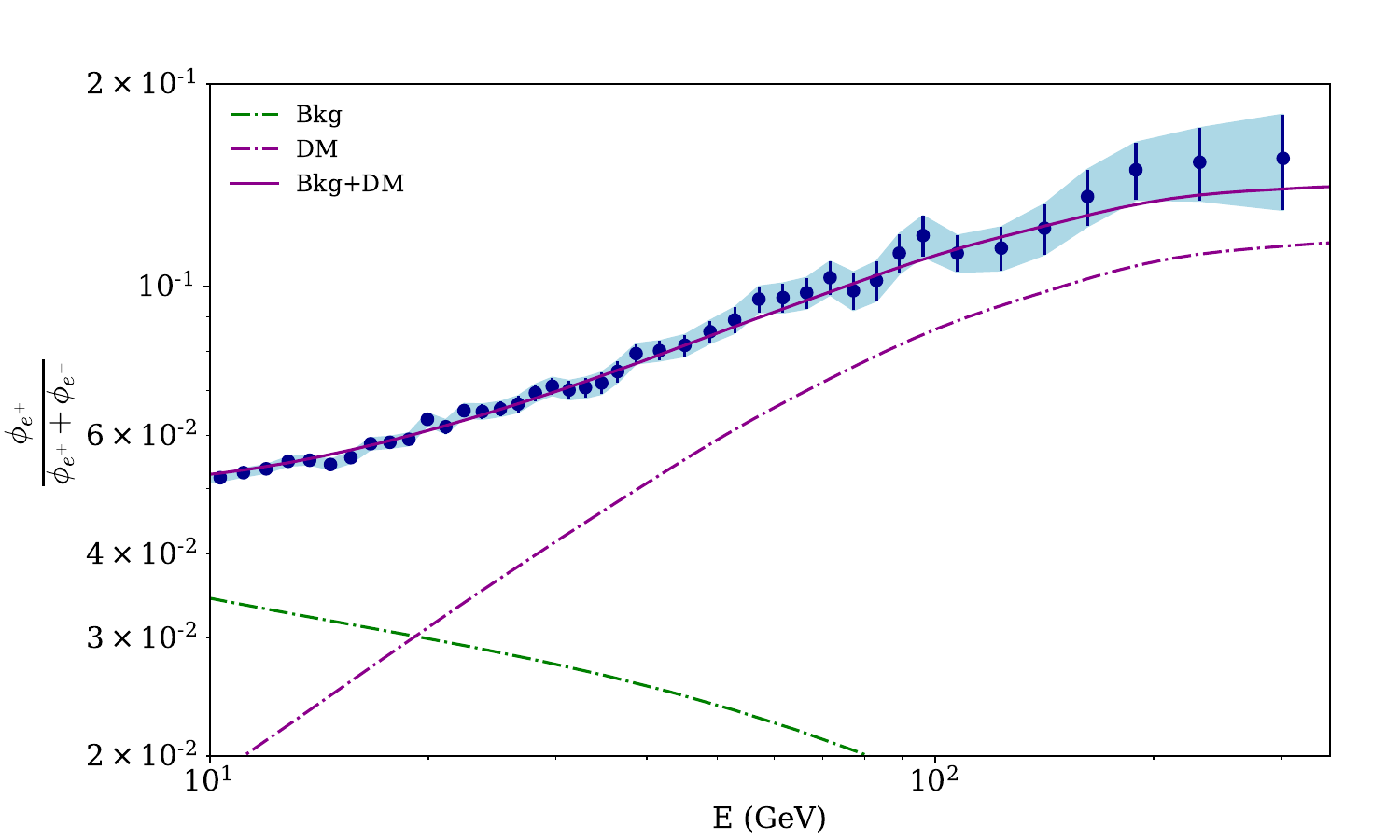}
\caption{
Methodology of exclusion for branons. 
The upper panel represents a fit where the  background plus pulsars contributions have been assumed to explain the whole AMS-02 signal (model A). As an example, in the plot, the obtained values for the branon are $\{f=4.99\, {\rm TeV},\, M=20.40\, {\rm TeV}\}$ ($\chi^2=56.94$, $\chi_{red}^2=1.42$). The procedure to constrain the DM model has consisted of discarding the branon masses for a tension of the brane unable to explain the AMS-02 experimental data within their error bars.  
Model B is showed in the middle panel. The procedure to constrain the DM model has consisted of discarding the branon masses that are above the experimental data. A $\chi^{2}$ analysis has been also performed and subsequent exclusion maps as presented in Figure \ref{Fig_exclusion} obtained. In the lower panel, model C, pulsars are assumed to be completely absent (model C). This scenario corresponds to the best fit of AMS-02 with DM over a background. The branon mass lies at $M=38.1\, {\rm TeV}$ ($\chi^2=37.15$, $\chi_{red}^2=0.94$) for the same tension $f=4.99\, {\rm TeV}$. For brane tensions $f$ greater than $15$ TeV the DM signal is practically unobservable for all the different cases we have studied in this work.} 
\label{Figure_XXX}
\end{figure}

The comparison with the AMS-02  experimental results has been done using the positron fraction from the 
CRDB database 
\cite{Maurin:2013lwa} where the data for the quantity
\begin{eqnarray}
{\cal F}_{e^+}=\frac{\Phi_{e^+}(E)}{\Phi_{e^+(E)}+\Phi_{e^-(E)}}
\label{positron_fraction}
\end{eqnarray}
have been provided.  
%
%
\begin{figure}[tbp]
\centering 
\includegraphics[width=.475\textwidth, height=6.0cm]{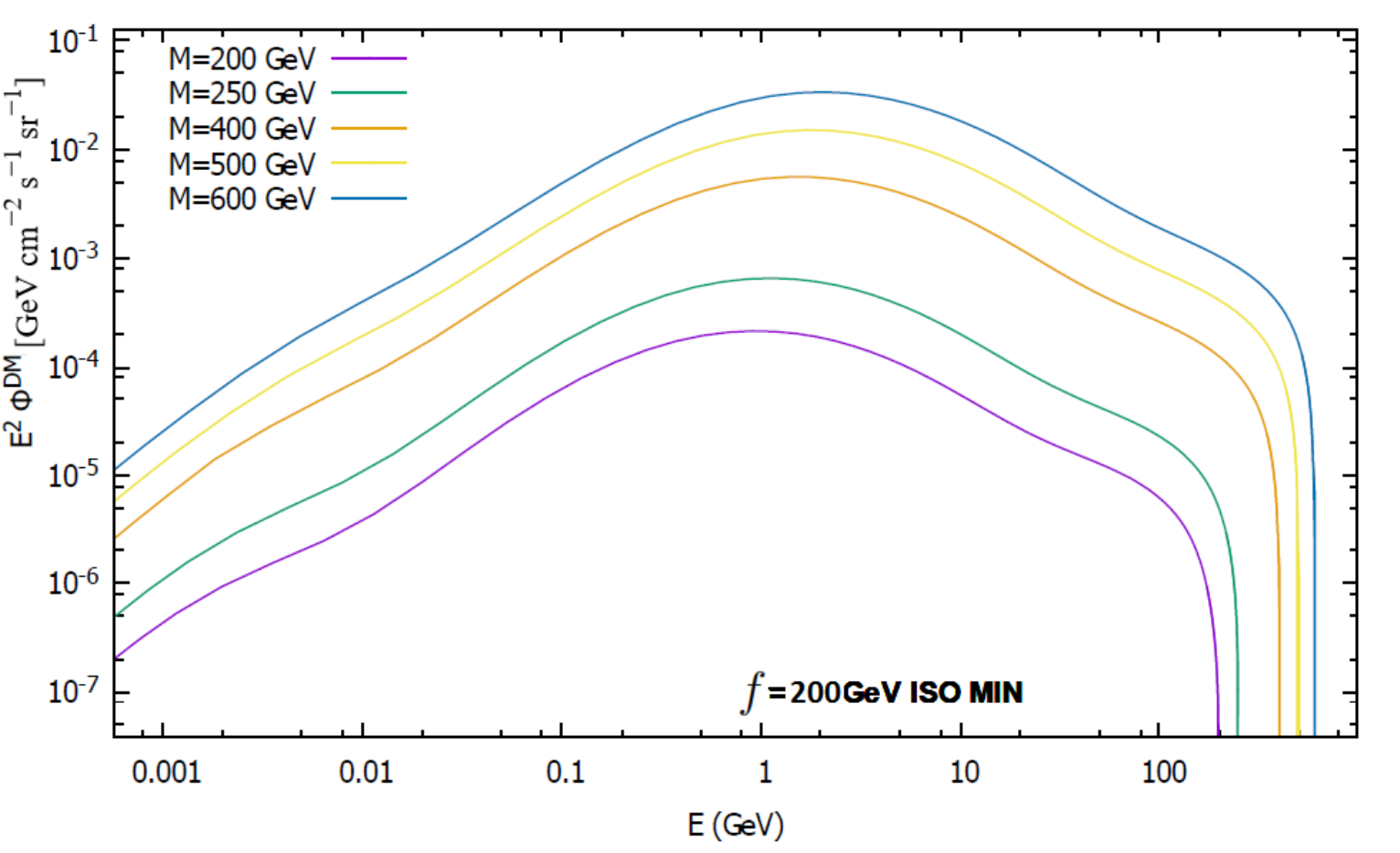}
\hfill
\includegraphics[width=.475\textwidth,height=6.0cm,origin=c]{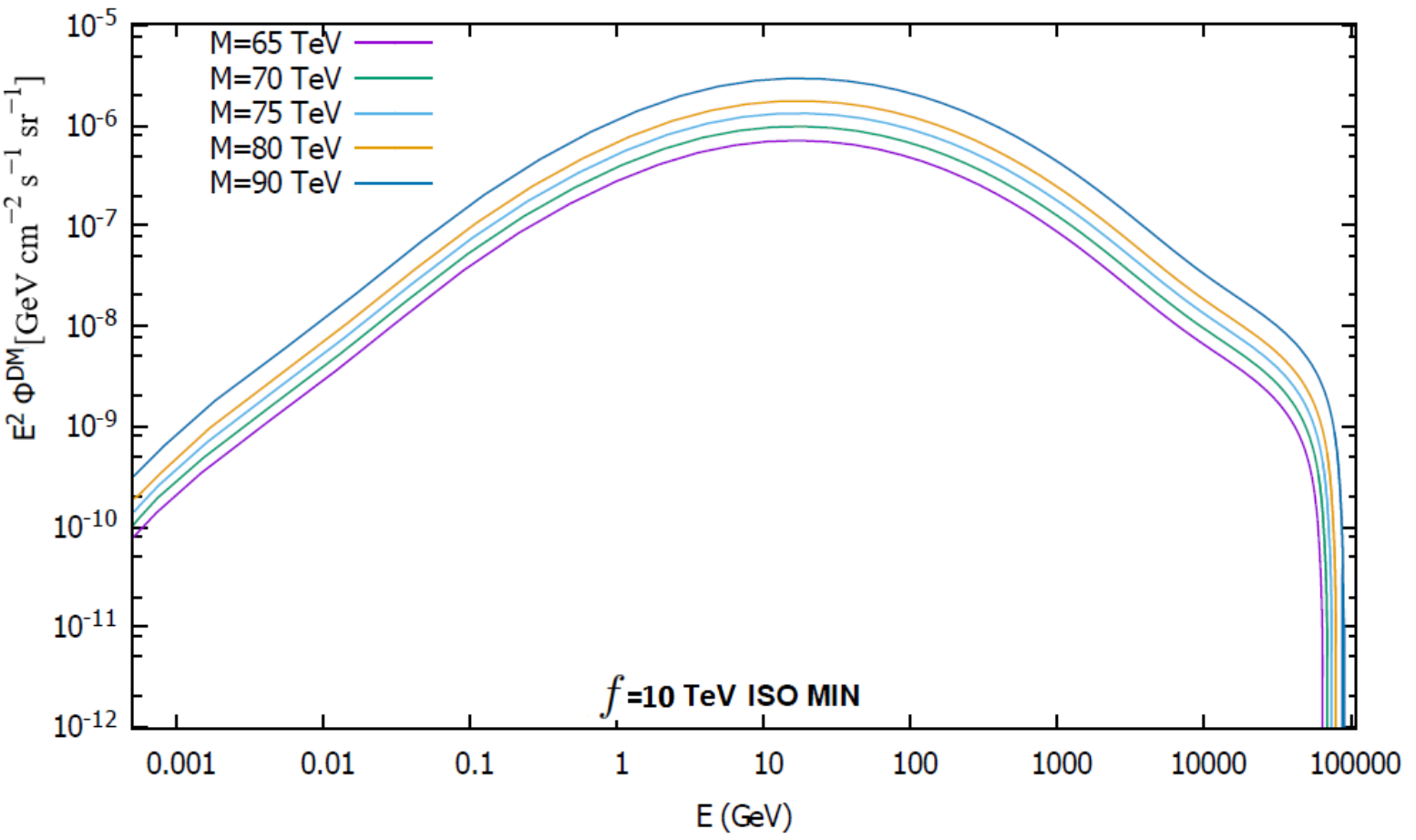}
\caption{Flux of positrons at the Earth after the propagation, for a minimal diffusion of a Isothermal profile and constant magnetic field equal to $6$ $\mu$G for different masses. Top panel corresponds to $f=200$ GeV whereas the bottom one corresponds to $f=10^4$ GeV. The characteristic cut off of the annihilation signature at the mass of the branon can be observed in the panels. 
}
\label{Fig_flux_positrons}
\end{figure}
In  our  analysis  we  shall  pursue  three reasoning  lines.  In  the  first  scenario  (Model  A),  AMS-02 data  can  be  perfectly  described  by  a  pulsar  contribution  added to  an  astrophysical  background.  Taking into account that the $e^+/e^-$ signal from branon annihilation depends on two parameters, namely $f$ and $M$, in this scenario we have put constraints on these
parameters aiming at the DM contribution being negligible
and consequently the background plus pulsars being able to explain the positron fraction data at $95\%$ confidence level. In a second scenario (Model B), we have added the branon signal to the background contribution without pulsars (green dashed line in Figure \ref{Figure_XXX}) given the fact that the pulsar signal would be assumed as a free-model contribution. In other words, in this analysis the pulsar signal needs to be understood as the remaining signal necessary to fit the AMS02 data in the event DM does not suffice to do so. This situation is described by the solid purple line in the middle panel in Figure \ref{Figure_XXX}, in which DM together with the no-pulsar astrophysical background is not enough to explain the data. As such, we have considered as viable the cases in which the total signal remains below the data and ruled out those cases exceeding the experimental data.



In order to describe the three scenarios above, and
according to the model in \cite{Graziani:2017fol,Aguilar:2014mma}, we divide the non-modulated (NM) positron flux $\Phi^{\rm 
NM}_{e^{+}}$ in its astrophysical (non-DM) primary (prim) and secondary (sec) components of cosmic rays to constrain the AMS-02 signal using branons, taking into account that
\begin{eqnarray} 
\Phi^{{\rm NM}}_{e^{+}}(E)=\Phi^{{\rm prim}}_{e^{+}}\left(E\right)+\Phi^{{\rm sec}}_{e^{+}}\left(E\right)+\Phi^{\rm{\rm{DM}}}\left(E\right)
\end{eqnarray} 
with
\begin{eqnarray}
\Phi^{{\rm prim}}_{e^+} (E)&=&C_{s}E^{-\gamma_{s}}\exp\left(-\frac{E}{E_{s}}\right),\nonumber\\
\Phi^{{\rm sec}}_{e^+} (E)&=&C_{e^+}E^{-\gamma_{e^{+}}},\nonumber
\end{eqnarray}
where both $\Phi^{{\rm prim}}_{e^+}$ and $\Phi^{{\rm sec}}_{e^+}$ refer to standard (non-DM) astrophysical 
contributions.
The contribution for the positron primary flux 
can be produced in a pulsar environment under strong magnetic fields through the decay of high-energy photons into positron-electron pairs. Secondary ones are produced in the primary component collisions with the ISM in the process of spallation.  
Consequently, and following the force field approximation \cite{gleeson1967cosmic, gleeson1968solar}
\begin{eqnarray}
\Phi_{e^+}(E)=\frac{E^{2}}{(E+\phi_{e^+})^{2}}\Phi^{{\rm NM}}_{e^+}\left(E+\phi_{e^+}\right).\
\label{Positron_flux}
\end{eqnarray}
In the case of electrons, the flux caused by nearby SNRs and pulsars, 
can be correctly fitted through a combination of two power laws \footnote{
The parameters of the fluxes $C_{s},\gamma_{s}$, $E_{s},C_{e^{+}}$, $\gamma_{e^{+}}$, $\phi_{e^{+}}$, $C_{e^{-}}$, $\gamma_{e^{-}}$, $\phi_{e^{-}}$, $\gamma_{1}$, $\gamma_{2}$ can be obtained from \cite{Ibarra:2013zia}. Such values are suitable for a range of energies between $2$ and $350$ GeV. 
} as follows
\begin{eqnarray}
\Phi_{e^{-}}(E)&=&\frac{E^2}{(E+\phi_{e^-})^2}\left[C_{1}(E+\phi_{e^{-}})^{\gamma_{1}}\right.\nonumber\\
&&+\left.C_{2}(E+\phi_{e^{-}})^{\gamma_{2}}\right]\,.
\label{Electron_flux}
\end{eqnarray}
Hence, expressions  (\ref{Positron_DM_flux}), (\ref{Positron_flux}) and (\ref{Electron_flux}) can be used to calculate the theoretical positron fraction ${\cal F}_{e^+}$ as in (\ref{positron_fraction}). In particular, 
the positron fraction can be calculated for the $(f, M_{\rm })$ parameter space describing DM particles originated in brane-world scenarios. For illustrative purposes, we have considered one extra dimension in a range of masses $M = 200\,{\rm GeV} - 100\,{\rm TeV}$. In this range, branons mainly annihilate via $Z\bar{Z}$ and 
$W^{+}W^{-}$ as seen in Figure \ref{Fig_BR_branons}. These bosons will then generate positrons. 
We have performed our calculations both for a model of minimal propagation (MIN) of a Isothermal profile, for a medium propagation (MED) of a NFW profile and  for a maximum propagation (MAX) of a NFW profile. The magnetic field taken was constant and  fixed to $B=6$ $\mu$G.

In Figure \ref{Fig_flux_positrons} we observe that the positron fraction increases with the branon mass $M$ 
due to the fact that the annihilation cross sections presented in  (\ref{cross_section_WZ}) also increase with the mass. This causes an enhancement in the source term $Q(\textbf{r},E)$ appearing Eq. (\ref{Q}). Then, the dependence of the cross section with $M$ compensates the $M^{-2}$ suppression which explicitely appears in  $Q(\textbf{r},E)$. 
This will result in the source term for the annihilating DM being inversely proportional to the square of the WIMP mass. Consequently the positron fraction would decrease with the DM mass. Notwithstanding, in the case of branons the cross section scales as $M^{6}$ for massive gauge fields as seen in Eq. (\ref{cross_section_WZ}), so the signal of positrons increases with the branon mass.

In addition, the effect of the brane tension is to suppress the probability of interaction between DM particles as seen once again in Eq. (\ref{cross_section_WZ}). Accordingly, a suppression of the positron fraction with $f$ can be seen in Figure \ref{Fig_exclusion}. 
However, for both low masses and tensions, the ${\cal F}_{e^+}$ fraction suppression caused by the brane tension
can be compensated for. For instance, 
for a tension of $f=200$ GeV, masses providing a measurable enhancement in the positron fraction lie around $M=200$ GeV. However, as tension $f$ gets bigger the observed excess in the positron fraction cannot be reproduced even for very high DM masses. For instance, for a tension $f=10$ TeV, 
branons with $M=50$ TeV would not produce any signature in the AMS-02 results. 

In Figure \ref{Figure_XXX} we show the pathway to constrain the signal from branons. For models A and B we have calculated
the positron fraction in a branon mass interval ranging from 200 GeV to 100 TeV. In principle, the brane tension has been left free in order to acquire a significant positron fraction in the range of masses above. We have found that Model A produces tighter constraints in the $(f,\, M)$ diagram as shown in Figure \ref{Fig_exclusion}. This is due to the fact that in this scenario the AMS-02 data are explained by astrophysical sources, i.e., pulsars, so the addition of a small amount of DM indeed spoils that explanation. As explained above, for Model B we have assumed a background without pulsar contributions to explain AMS-02 data. It turns out that DM contribution needed to exceed the experimental data is bigger than in the previous Model A. 


\section{\label{Sec:VI}6. Conclusions}

\begin{figure}[htbp] 
    \centering
 \includegraphics[width=.475\textwidth,height=6.0cm]{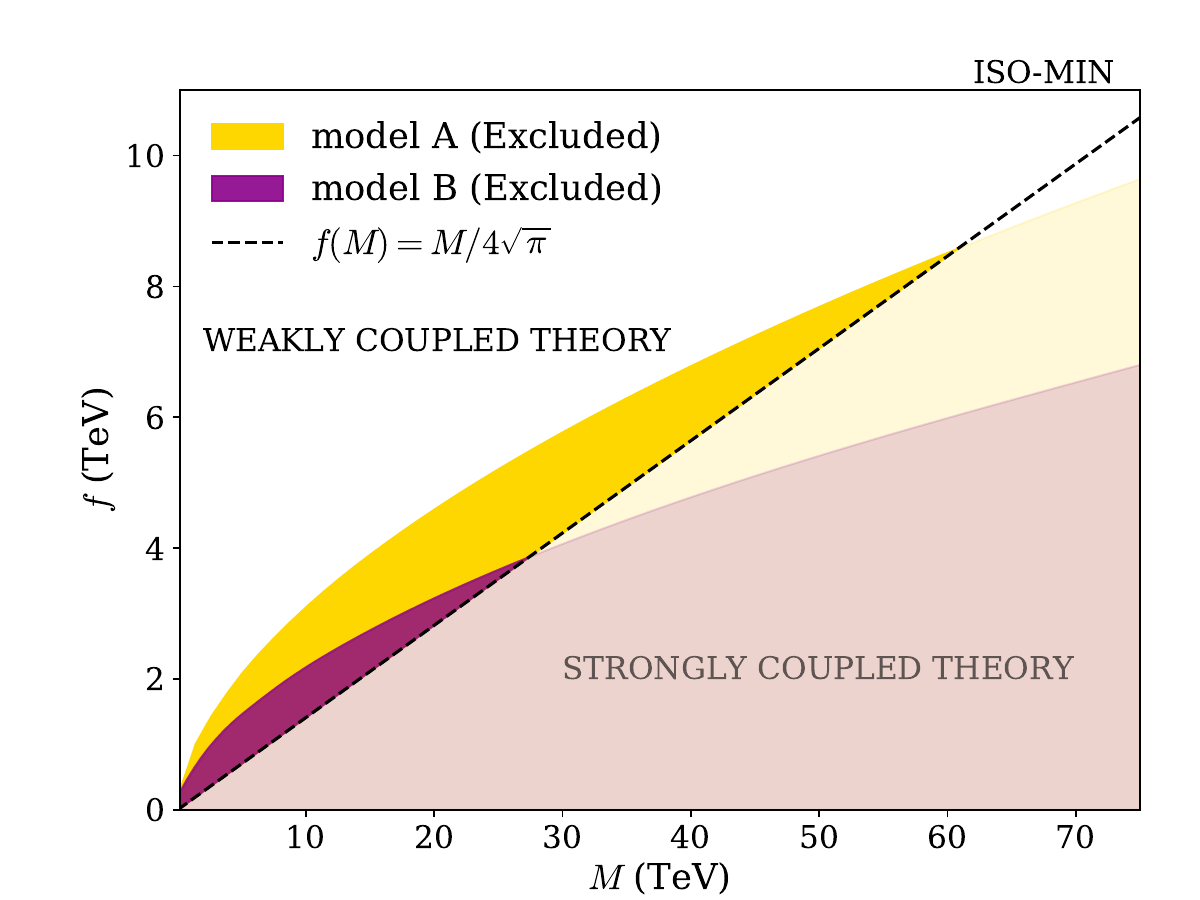}
 \includegraphics[width=.475\textwidth,height=6.0cm]{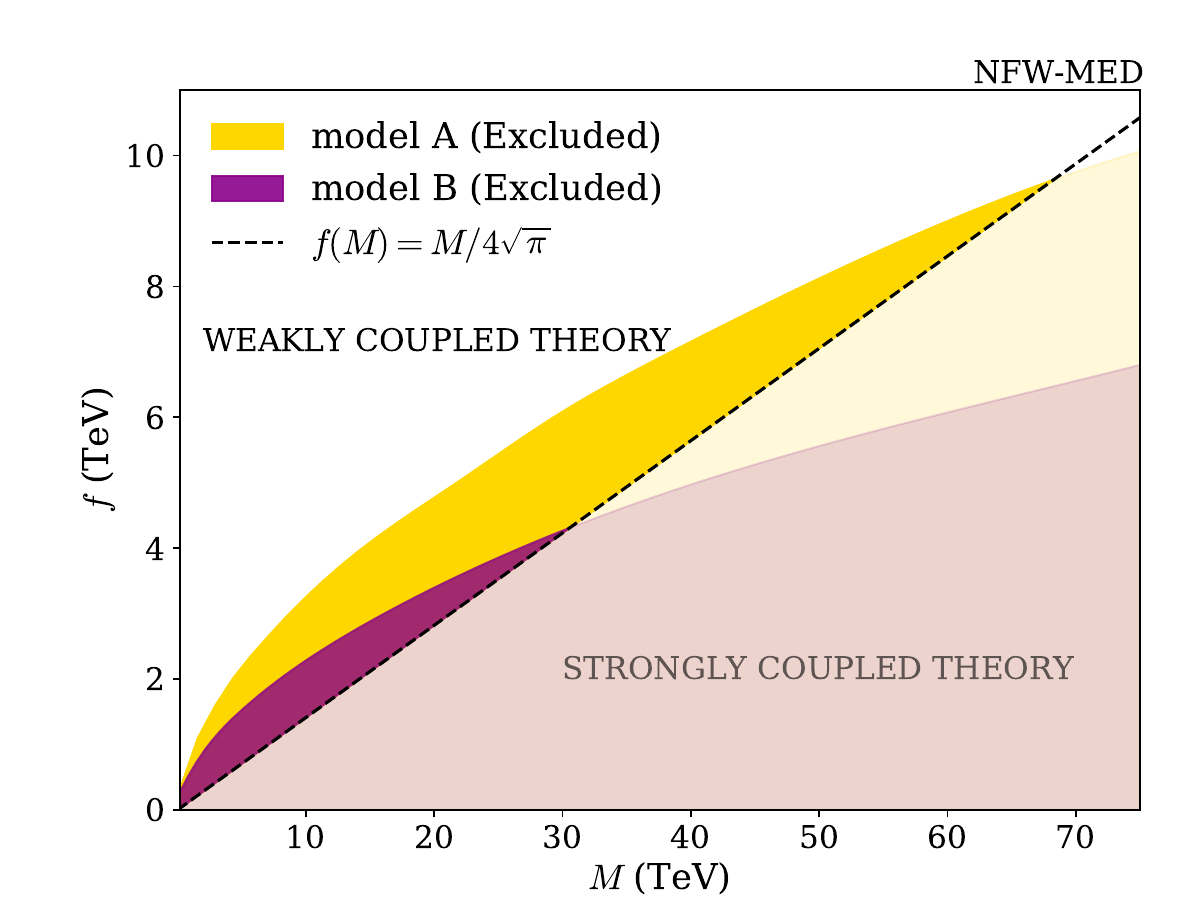}
\includegraphics[width=.475\textwidth,height=6.0cm]{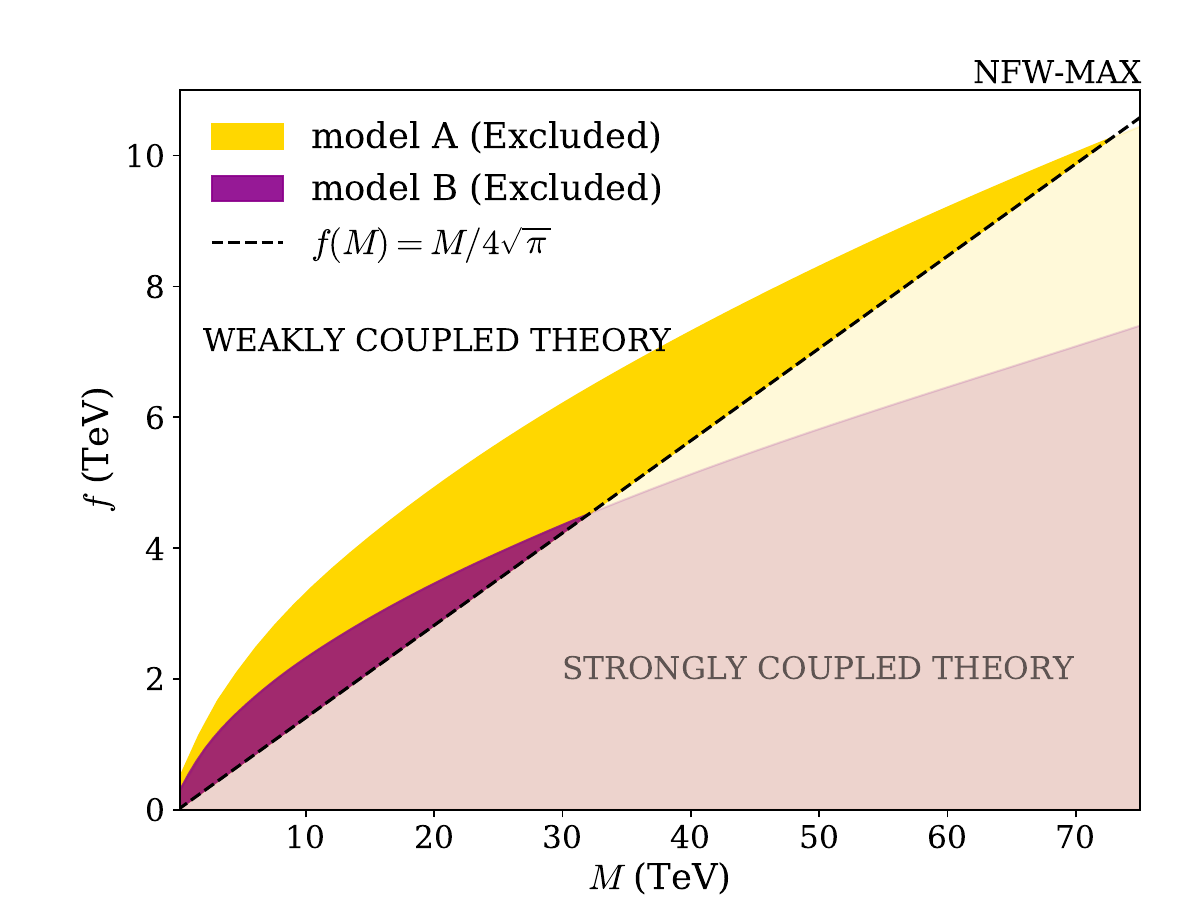}
\caption{Exclusion diagrams for the  model A (purple) and model B (yellow). Bright coloured areas indicate the excluded region in the parameter space $(f,\, M)$ assuming the 
ISO-min (left panel), NFW-med (mid panel) and NFW-max (right panel) dark matter profiles. The lowest mass in this study is $200\, {\rm GeV}$, ensuring the annihilation into the channels $W$,$Z$, $t$. Dashed black line sets the limit of the perturbative theory for branons and the validity of tree-level versus loop branon effects \cite{Cembranos:2005jc}. The exclusion is only valid when the coloured area lies above the line (weakly coupled region).}
\label{Fig_exclusion}
\end{figure}

In this work we have set constraints on the parameter space for brane-world theories using the AMS-02 positron excess. In order to do so, we have assumed three different pathways; one in which the positron fraction can be fully explained within the error bars 
by different astrophysical sources so the eventual dark matter contribution needs to remain negligible within the statistical significance. A second one in which dark matter contribution plus background cannot exceed the experimental results.  The third pathway has consisted of finding the model that best fit the data without other astrophysical sources.  This approach has been followed in a series of recent studies for a wide variety of dark matter models 
\cite{Feng:2017tnz,Laletin:2017qei,Cheng:2016slx,Carquin:2015uma,DiMauro:2015jxa, Lin:2014vja,Lu:2015fdn, Ibarra:2013zia}.
%
Since brane-world theories provide natural candidates for dark matter particles, 
we have generated exclusion diagrams for the space $(f,\,M)$ using the models A and B described in the Section 5\ref{Sec:V}. In the parameter
space, $f$ is the brane tension and $M$ the mass of the dark matter particle identified with the brane oscillations, dubbed
branons, which indeed satisfy usual properties of standard WIMPs. Figure \ref{Fig_exclusion} encapsulates such exclusion diagrams for one extra dimension. In order to determine  excluded regions in the $(f,\,M)$ space,  we have used a likelihood analysis at the 95 \% confidence level with $40$ degrees of freedom (d.o.f), although we are aware that the $\chi^{2}$ test is not precisely correct, since the AMS-02 error bars do not follow a Gaussian distribution to a tee. Thus, strictly speaking, it would be necessary 
to perform variations in the whole space of both astrophysical and dark matter model parameters in order to obtain the statistical significance of the predictions. Nonetheless, as far as the level of accuracy we are looking for, our approximation is sufficient to put constraints on branons features.

The dark matter source term as provided by Eq. (\ref{Q}) shows that the positron fluxes from branons (annihilating in all the channels) can be computed as a linear superposition of the flux of every channel separately, so the  diffusion equation can be solved for every channel.
%
Also, as can be seen from the branon cross sections expressions 
(\ref{cross_section_fermion})-
(\ref{cross_section_complex}),
the brane tension will appear as a multiplicative factor $f^{-8}$ in (\ref{Q}) for every channel. Consequently $f^{-8}$ could be factorised out. 
%
 It would then be possible to include the usual astrophysical boost factor 
which accounts for
the clumps of dark matter in the halo as a kind of effective tension.
%
Since the branon cross section scales as $f^{-8}$, the source term in  (\ref{Q}) is highly dependent on the tension $f$. Consequently, the eventual inclusion of a multiplicative boost factor would not be significant taking into account its expected value \cite{Anderhalden:2013wd}. 

The $(f,\,M)$ exclusion area as obtained in this study using AMS-02 data has been compared 
with previous analyses \cite{Cembranos:2016jun} extending the exclusion area constrained by colliders and supernovae catalogues.  
Our analysis has shown that both the chosen dark matter density profiles and different diffusion models do not significantly modify the parameter space $(f,\, M)$ exclusion limits, as can be seen throughout Figure \ref{Fig_exclusion} panels. Our results show that the Navarro-Frenk-White dark matter profile for both medium and maximum diffusion provide the tightest constraints in comparison with the Isothermal profile.  
 The first two involve a higher amount of dark matter at the centre of the galaxy and a bigger $\lambda_{D}$, so that a bigger fraction of products than expected in the Isothermal minimum model would make it to the Earth. The limit in which the description of the effective theory for branons is valid (the weakly coupled region) is also represented in Figure \ref{Fig_exclusion}. This limit characterises how strongly coupled the brane is and the validity of the tree-level with respect to the loop branon effects \cite{Cembranos:2005jc}.   
Indeed, the dark matter models computed in this study have a thermally averaged cross section greater than $ \langle \sigma v\rangle =  3\cdot10^{-26} \text{cm}^{3}/\text{s}$. In other words,  our analysis would not able to rule out the thermal region in which branons acquire the required dark matter abundance by the standard freeze-out mechanism.
However, as we have mentioned in this section, the addition of a boost factor could slightly modify the parameters space constraints plotted in Figure \ref{Fig_exclusion} and reduce the viable region of the parameters space.

In fact, the most notable difference in the exclusion limits when comparing different models of diffusion and dark matter density profiles turns out to be between the Isothermal minimal and the Navarro-Frenk-White maximum profiles. However, this difference is not very significative; comparing the exclusion diagrams in Figure \ref{Fig_exclusion} we observe that the difference between the edges of the exclusion area for both cases differ less than $12$ TeV for masses and $2$ TeV for tensions when in fact our study covers the large range of $98$ TeV in masses and $10$ TeV in tensions. Thus, it seems that only the tension of the brane (and the mass but only in the range of low masses), plays a crucial role in the process of injecting positrons to the medium, and, hence in the exclusion diagram.

Concerning the third scenario where there is no contribution of pulsars, we have performed a $\chi^2$-analysis in the $(f,\,M)$ parameter space. For the sake of simplicity, we have illustrated this scenario for a Navarro-Frenk-White profile with maximum diffusion. Herein we have found a global minimum at $M=38.1 \pm 0.2$ TeV and $f=4.99 \pm 0.04$ TeV  ($\chi^2=37.15$, $\chi_{red}^2=0.94$). In fact this scenario seems to be in agreement with recent HAWC detector measurements \cite{Abeysekara:2017old} for AMS-02 data, showing that this scenario where there is no contribution of pulsars, may be the most suitable of all the studied in the bulk of the paper. Indeed, HAWC measurements have provided tight constraints for the positrons diffusion in pulsars Geminga and PSR B0656+14 showing that the injection of particles in the surroundings of these sources is not really energetic to reach the Earth.

Finally, we allude to several other studies, separate to this one, which can be done with AMS-02 measurements. Indeed, both the total flux for electrons and positrons have been separately measured with AMS-02 and separate analysis of every flux might constrain indirect dark matter signals as well \cite{Carquin:2015uma}. However, as has been described in the bulk of our study, we have come across a high sensitivity of the positron fraction with the brane tension, so further analysis with AMS-02 results does not seem essential to obtain competitive constraints in the branon parameters space.

%

\section{Acknowledgements}
This work was partly supported by the projects FIS2014-52837-P (Spanish MINECO) and FIS2016-78859-P (AEI / FEDER, UE), and Consolider-Ingenio MULTIDARK CSD2009-00064.
AdlCD acknowledges financial support from projects FPA2014-53375-C2-1-P Spanish Ministry of Economy and Science, 
FIS2016-78859-P European Regional Development Fund and Spanish Research Agency (AEI), 
CA15117 CANTATA and CA16104 COST Actions EU Framework Programme Horizon 2020,  
CSIC I-LINK1019 Project, Spanish Ministry of Economy and Science,
University of Cape Town Launching Grants Programme and National Research Foundation grants 99077 2016-2018 (Ref. No. CSUR150628121624), 110966 Ref. No. BS1705-09230233 and the NRF Incentive Funding for Rated Researchers, Ref. No. IFR170131220846. AdlCD would like to thank the 
Abdus  Salam  International  Centre  for  Theoretical  Physics  (ICTP, Trieste) for its hospitality in the latest stages of the manuscript. 
PKSD acknowledges NRF for financial support.
MMI thanks Carlos Mu\~noz (IFT/CSIC-UAM), Roy Maartens (UWC) and Roberto Lineros (IFIC) for useful discussions and acknowledges financial support from the University of Cape Town Doctoral Fellowships and the Erasmus+ Alliance-4-Universities Mobility Programme.

\end{document}